%% file: sample.tex
\documentclass[sigconf]{aamas} 
\usepackage{balance} % for balancing columns on the final page
\title[Multi-Agent RL for Demand Response]{Multi-Agent Reinforcement Learning for Fast-Timescale Demand Response of Residential Loads}

\author{Vincent Mai}
\affiliation{
  \institution{Robotics \& Embodied AI Lab, Mila}
  \city{Université de Montréal}
  \country{Canada}}
\email{vincent.mai@umontreal.ca}

\author{Philippe Maisonneuve}
\affiliation{
  \institution{GERAD \& Mila}
  \city{Polytechnique Montréal}
  \country{Canada}}
\email{philippe.maisonneuve@polymtl.ca}

\author{Tianyu Zhang}
\affiliation{
  \institution{Mila}
  \city{Université de Montréal}
  \country{Canada}}
\email{tianyu.zhang@mila.quebec}

\author{Hadi Nekoei}
\affiliation{
  \institution{Mila}
  \city{Université de Montréal}
  \country{Canada}}
\email{nekoeihe@mila.quebec}

\author{Liam Paull}
\affiliation{
  \institution{Robotics \& Embodied AI Lab, Mila}
  \city{Université de Montréal}
  \country{Canada}}
\email{liam.paull@umontreal.ca}

\author{Antoine Lesage-Landry}
\affiliation{
  \institution{GERAD \& Mila}
  \city{Polytechnique Montréal}
  \country{Canada}}
\email{antoine.lesage-landry@polymtl.ca}

\begin{abstract}
To integrate high amounts of renewable energy resources, electrical power grids must be able to cope with high amplitude, fast timescale variations in power generation. Frequency regulation through demand response has the potential to coordinate temporally flexible loads, such as air conditioners, to counteract these variations. Existing approaches for discrete control with dynamic constraints struggle to provide satisfactory performance for fast timescale action selection with hundreds of agents. We propose a decentralized agent trained with multi-agent proximal policy optimization with localized communication. We explore two communication frameworks: hand-engineered, or learned through targeted multi-agent communication. The resulting policies perform well and robustly for frequency regulation, and scale seamlessly to arbitrary numbers of houses for constant processing times.%, where classical methods fail.
\end{abstract}


\usepackage{gensymb}
\usepackage{algorithmic}
\usepackage{graphicx}
\usepackage{textcomp}
\usepackage{xcolor}
\usepackage{url}
\usepackage{multirow}
\usepackage{caption}
\usepackage{subcaption}
\usepackage[utf8]{inputenc}
\usepackage[T1]{fontenc}
\usepackage{multirow}
\usepackage{chngpage}
\usepackage{adjustbox}
\usepackage{hyperref}
\usepackage{mathtools}

\newcommand{\BibTeX}{\rm B\kern-.05em{\sc i\kern-.025em b}\kern-.08em\TeX}

\begin{document}

\pagestyle{fancy}
\fancyhead{}
\maketitle 
%%%%%%%%%%%%%%%%%%%%%%%%%%%%%%%%%%%%%%%%%%%%%%%%%%%%%%%%%%%%%%%%%%%%%%%%

\input{Contents/1_Introduction}
\input{Contents/2_RelatedWork}
\input{Contents/3_ProblemFormulation}

\input{Contents/4_Methods}
\input{Contents/5_Results}

\input{Contents/6_Conclusion}

%%%%%%%%%%%%%%%%%%%%%%%%%%%%%%%%%%%%%%%%%%%%%%%%%%%%%%%%%%%%%%%%%%%%%%%%

\clearpage
\bibliographystyle{ACM-Reference-Format} 
\bibliography{biblio.bib}

%%%%%%%%%%%%%%%%%%%%%%%%%%%%%%%%%%%%%%%%%%%%%%%%%%%%%%%%%%%%%%%%%%%%%%%%
\clearpage

\appendix
\input{Contents/Appendix/A_Notation}
\input{Contents/Appendix/B_CarbonAccounting}
\input{Contents/Appendix/C_EnvironmentDetails}
\input{Contents/Appendix/D_AlgorithmsDetails}

\input{Contents/Appendix/E_Proofs}

\end{document}

%% file: Contents/1_Introduction.tex
\section{Introduction}
\label{sec:introduction}

To achieve the United Nations' climate change target of limiting global warming at +1.5$\degree$C, global electricity generation must transition from fossil fuels to renewable energy sources such as wind turbines and solar panels. 
In 2019, according to the International Energy Agency, electricity and heat production accounted for 40$\%$ of global emissions \cite{IEA_Emissions}, as 64\% of it is generated from burning fossil fuel \cite{IEA_EnergySources}.
The electricity sector must thus move from a conventional, fuel-burning paradigm to a renewable, natural phenomenon-based generation, e.g., wind turbines and solar photovoltaics. 
Renewable energy generation is subject to short-term, high-amplitude variations, referred to as intermittency. As an example, a cloud passing will lead to a sudden drop in the solar-based generation, followed by a sharp increase when the sky becomes clear again. These changes can happen at the scale of a few seconds, and create major challenges for power grid operators: to ensure the stability of the electric grid, a near-perfect balance between the power demand and the generation is critical~\cite{kundur2007power}. In other words, power generation and consumption must be equal at all times. Hence, trading a constant, deterministic generation for an intermittent, uncertain one exacerbates the need for power balancing. At the second timescale, this balancing task is referred to as frequency regulation~\cite{bevrani2010renewable,taylor2016power}.

On the power generation side, solutions such as excess energy storage in batteries or support with fossil fuel plants require large investments and are not renewable respectively.
Alternatively, demand response programs~\cite{siano2014demand} can be introduced to mitigate renewable intermittency~\cite{taylor2016power}.
The demand response approach aims at adjusting the power demand to meet the supply by coordinating loads temporally. These loads must be flexible, i.e., capable of modulating their consumption while fulfilling their own purpose. This does not apply to, for example, computer monitors, which must be fully powered when they are in use. Thermostatic loads, such as heating, air conditioning or water heaters, are instead ideal candidates: they do not need to be turned on at all times, as long as the temperature of the air/water is within the user's preference range \cite{callaway2009tapping}. They are also widely deployed and they represent a significant part of global power consumption~\cite{IEA_Cooling,mathieu2012state}. The frequency regulation objective differs from peak-shaving, for which the objective is load shifting over, e.g., a day. It is instead to leverage the loads' flexibility to balance out high-frequency variations in power generation.

In this paper, we focus on the task of fast timescale demand response for frequency regulation using \textit{residential air conditioners}. This presents several physical and algorithmic constraints:
(1) air conditioners are \textbf{discretely} powered, i.e., \textsc{on} or \textsc{off}, which limits the control flexibility; (2) they are subject to hardware \textbf{dynamic constraints} such as lockout: once turned \textsc{off}, they must wait some time before being allowed to turn back  \textsc{on} to protect the compressor; (3) as the context is residential, privacy is important and \textbf{communications should be limited}; (4) to provide enough power flexibility to the grid, a large aggregation of loads must be considered: the method must be \textbf{scalable}; (5) for easier implementation, the control should also be \textbf{decentralized} with \textbf{localized communications} ; (6) the decisions must be taken at a \textbf{few seconds timescale}; and finally (7) the control algorithm should be able to \textbf{cope with uncertainty} in the future regulation signal. 

These constraints impede the deployment of classical methods. Greedy algorithms are centralized and have difficulty accounting for long-term dynamic constraints \cite{lesage2021online}. Standard model predictive control is also centralized, and even decentralized versions solve a multi-period combinatorial optimization problem that does not scale with the number of agents \cite{Jin, Chen_Francis_Pritoni_Kar_Bergés_2020}. 
We propose to tackle this problem by using multi-agent reinforcement learning (MARL) to learn decentralized and scalable policies (4) with discrete and constrained control (1, 2) and limited and localized communications (3, 5). Once learned, these policies can take the best decisions in real time (6) based on expected value over uncertainty (7). As this problem combines the most important current challenges of MARL, i.e., communication, long-term credit assignment, coordination, and scalability~\cite{Gronauer_Diepold_2021}, it is also an interesting benchmark for MARL algorithms.
We train our agents with Multi-Agent Proximal Policy Optimization (MA-PPO) \cite{yu2021surprising} with Centralized Training, Decentralized Execution (CTDE). \cite{Kraemer_Banerjee_2016}. Two local communication frameworks are tested -- hand-engineered and learned -- and both outperform the baselines.
Our main contributions are threefold:
\begin{itemize}
    \item an open source, multi-agent environment\footnote{The code is hosted on \url{https://github.com/ALLabMTL/MARL_for_fast_timescale_DR}.} simulating the real-world problem of frequency regulation through demand response at the second timescale. The simulator is compatible with the OpenAI Gym \cite{OpenAIGym} framework.
    \item two decentralized, fast-responding agents\footnotemark[1] trained by MA-PPO. The first one has a hand-engineered communication strategy, while the second one learns what data to share through Targeted Multi-Agent Communication (TarMAC) \cite{Das_2019}. Both outperform baselines on two-day simulations.
    \item an in-depth analysis of the dynamics, communications, scalability and robustness of the trained agents.
\end{itemize}

In the next section, we describe prior work in the field of demand response and MARL. In Section~\ref{sec:problem}, we describe the environment and formulate the problem. The classical and learning-based methods are described in Section~\ref{sec:methods}. Finally, Section~\ref{sec:results} presents the experimental results and analyses of the agents' performance, dynamics, robustness, and scalability. 

%% file: Contents/2_RelatedWork.tex
\section{Related Works}
\label{sec:relatedworks}

Frequency regulation through demand response is commonly tackled by model predictive 
control (MPC) \cite{Wu,lee2015optimal,Olama,Jin, maasoumy2014model, mathieu2012state}, where the best action is chosen based on trajectory prediction over a given horizon, sometimes combined with machine learning \cite{Dusparic, LAURO2015494, Ahmadiahangar}. 
Apart from \cite{liu2015model}, these works do not consider short-term dynamic constraints such as lockout. MPC approaches rely on mixed-integer programming, which does not scale sustainably with higher numbers of agents, preventing control at fast timescales. Moreover, these works generally require a centralized entity to access residences' data, leading to confidentiality issues.
An alternative method of multipliers-based distributed MPC approach was proposed in~\cite{chen2020cohort}. This approach did not consider the lockout constraint and is not compatible with fast timescale decision-making as it requires multiple centralized communication rounds at each time step in addition to solving several optimization problems and converting continuous setpoints to binary actions.

To tackle these problems, online optimization (OO) approaches~\cite{lesage2018setpoint, zhou2019online} have been used because of their high computational efficiency and scalability. In particular, \cite{lesage2021online} deploys OO for frequency regulation with binary control settings as is the case for ACs. However, these methods rely on greedy optimization and their lack of foresight leads to limited performance when facing dynamic constraints.
Reinforcement learning (RL) methods have been developed to address the longer timescale power balance problems such as peak shaving through demand response \cite{Aladdin} or coordination of loads and generators \cite{Roesch, Yang_Hao_Zheng_Hao_Fu_2019}. The CityLearn environment \cite{cityLearn} proposes a standard environment for multi-agent RL (MARL) for demand response, upon which are developed methods such as \cite{Piggot} to regulate the voltage magnitude in distribution networks using smart inverters and intelligent energy storage management, and \cite{Canteli_Jose} for load shaping of grid-interactive connected buildings. The AlphaBuilding ResCommunity environment~\cite{wang2021alphabuilding} then implements detailed thermal models. Both CityLearn and AlphaBuilding ResCommunity, however, consider longer timescale control, which makes them inadequate for high-frequency regulation and removes the ACs' lockout and binary constraints. 
The PowerGridworld \cite{Pwgworld} environment, a more flexible alternative to CityLearn, allows fast-timescale simulation but does not provide a detailed thermal model of loads, options for lockout or binary control, or classical baseline approaches to compare with.  
High-frequency regulation has been addressed by MARL, but only on the power generation side \cite{LEI}. We are unaware of any example in the literature deploying MARL for frequency regulation with demand response, with second-timescale control and flexible binary loads such as ACs which are subject to hardware dynamic constraints like a lockout. 

More generally, MARL has been developed for collaboration both in virtual environments such as Dota 2 \cite{OpenAIFive_2019}, Hide and Seek \cite{HideandSeek_2020} or Hanabi \cite{Hanabi_2021}, and in real-world environments such as traffic light control \cite{CoLight_2019}, single-house energy management \cite{Ahrarinouri_Rastegar_Seifi_2021} or ride-sharing \cite{Qin_Zhu_Ye_2022}. MARL problems pose several additional challenges to the RL settings \cite{Gronauer_Diepold_2021}, such as the non-stationarity of the environment, the need to learn coordination and communication, or the scaling of the training and deployment. Multi-agent adaptations of known RL algorithms, such as online PPO \cite{Schulman_Wolski_Dhariwal_Radford_Klimov_2017, yu2021surprising}, or offline DDPG \cite{DDPG_2019, MADDPG} and DQN \cite{Mnih_DQN}, have led to strong performance in many problems. However, some particular problems, such as the ones requiring communication with large numbers of agents, need specialized algorithms \cite{Jiang_Lu_2018}. TarMAC \cite{Das_2019}, for example, uses an attention mechanism to aggregate messages based on their importance.

%% file: Contents/3_ProblemFormulation.tex
\section{Problem Formulation}
\label{sec:problem}

\subsection{Environment}
\label{sec:environment}

The environment is a simulation of an aggregation of $N$ houses, each equipped with a single air conditioning (AC) unit. 
The outdoor temperature $T_{o,t}$ is assumed to be the same for every house, i.e., they are co-located in the same geographical region, and is simulated as sinusoidal with a one-day period.  Unless otherwise specified, the maximal temperature of 34 $\degree$C is reached at 6 pm and the minimal temperature of 28 $\degree$C at 6 am. $T_{o,t}$ is thus always above the target indoor temperature $T_T$ of 20 $\degree$C, so that every household can offer its flexibility to the grid. 
The environment model is updated every 4 seconds. Thermostatic loads modeled as multi-zone units and equipped with more than a single AC~\cite{amin2020optimal} is a topic for future work. More details about the environment are given in Appendix~\ref{sec:app_environmentdetails}. A notation table is provided in Appendix~\ref{sec:app_notation}.

\subsubsection{House thermal model}
Each house $i=1,2,\ldots, N$ is simulated using a second-order model based on Gridlab-D's Residential module user's guide \cite{GridLabD}. 
At time $t$, the indoor air temperature $T^i_{h, t}$ and the mass temperature $T^i_{m, t}$ are updated given the house characteristics $\theta^i_T$ (wall conductance $U^i_h$, thermal mass $C^i_m$, air thermal mass $C^i_h$ and mass surface conductance $H^i_m$), the outdoor temperature $T_{o, t}$, and the heat $Q^i_{a,t}$ removed by the AC. By default, the thermal characteristics are the same for each house and model a 100 square meter, 1-floor house with standard isolation. 
During training and deployment, the initial mass and air temperatures are set by adding a positive random noise over the target temperature. Although it is not used by default, the solar gain $Q_{s,t}$ can also be added to the simulation, as seen in Appendix~\ref{sec:app_solargain}.

\subsubsection{Air conditioners}
Once again based on Gridlab-D's guide \cite{GridLabD}, air conditioner $i$'s heat removal capacity $Q^i_{a,t}$ and power consumption $P^i_{a,t}$ are simulated based on the AC characteristics $\theta^i_a$, which include their cooling capacity $K^i_a$, their coefficient of performance $COP^i_a$ and the latent cooling fraction $L^i_a$. The model and parameters are also described in Appendix~\ref{sec:app_air_conditioner_model}. 
Additionally, a hard dynamic constraint is set to protect the compressor: after being turned \textsc{off}, it needs to wait a given amount of time before being allowed to turn \textsc{on} again~\cite{zhang2013aggregated}. This constraint is referred to as the lockout. By default, the lockout duration $l^i_\mathrm{max}$ is set to 40 seconds.

 %We remark that this differs from the total power consumption because this would also include the baseload consumption of the loads, i.e., the non-controllable power demand. %In this work, we are only interested in controlling the flexible power consumption and, for example, we can assume that the regulation signal accounts for the baseload without loss of generality.

\subsubsection{Regulation signal}
\label{sec:reg_signal}

The power system operator sends to the aggregator a signal $\rho_t$, which covers the complete aggregated load consumption: the systems we cannot control such as computers, washing machines, or lights, and the flexible power consumption, in our case, the ACs. Let, $\rho_t = D_{o,t}+s_t$ where $D_{o,t}$ is the power demand for the non-controllable loads and $s_t$ is the objective aggregated AC power consumption, i.e., the flexible load. We define $D_{a,t}$ as the power needed by the ACs to satisfy their thermal objectives, i.e., to keep the temperature around the target. To focus on the high-frequency variations of the power generation, we assume that $s_t$ is well behaved at low frequencies, i.e., its mean in the 5 minutes scale is $D_{a,t}$. A $0$-mean, high-frequency variation $\delta_{s, t}$ is added to represent renewable intermittency the aggregator wants to mitigate. We model the regulation signal as $s_t = D_{a,t} + \delta_{s,t}$. 

The aggregation flexible power consumption is the sum of all of the ACs' consumption: $P_t = \sum_i^N P^i_{a,t}$. The objective is to coordinate the ACs in the aggregation so that $P_t$ tracks $s_t$.

\paragraph{Base signal.} 
To compute the average needed power $D_{a,t}$, we created a dataset of the average power needed over a 5-minute period by a bang-bang controller without lockout -- which is optimal for temperature -- for all combinations of discrete sets of the relevant parameters.
At each time step, we interpolate the average power demand of each AC from this dataset and sum them to compute $D_{a,t}$. 
In practice, the base signal would be estimated or obtained from historical data. The aggregator would then consider its value when committing to track a signal $s_t$. This ensures that the required power adjustment is enough to maintain the houses at acceptable temperatures while providing flexibility to the grid.

\paragraph{Modelling high-frequency variations.}
The high-frequency variation $\delta_{s,t}$ is modelled with 1-D Perlin noise \cite{PerlinNoise}, a smooth, procedurally generated 0-mean noise. 
The Perlin noise produces $\delta_{p,t} \in [-1, 1]$, and we have  $\delta_{s,t} = D_{a,t} \beta_{p} \delta_{p,t}$ where $\beta_{p}$ is an amplitude parameter set to 0.9. Our Perlin noise is defined by 5 octaves and 5 octave steps per period of 400 seconds; it thus is the sum of noises with periods of 80, 40, 20, 10 and 5 seconds. More details are given in Appendix~\ref{sec:app_perlin}.

\subsubsection{Communication between agents}
To achieve coordination between agents, they must be able to communicate. For the agent implementation to be decentralized, flexible, and privacy-preserving, we consider limited and localized communications. This enables, for example, devices communicating with simple radio-frequency emitters, without the need for any further infrastructure. As such, we limit the communication to a number $N_c$ of neighbours. This is in line with the low-deployment investment argument for using demand response for frequency regulation.

\subsection{Decentralized Partially Observable Markov Decision Process}
\label{sec:dec-pomdp}
In this section, we formalize the above environment as a decentralized, partially observable Markov decision process (Dec-POMDP) characterized by the tuple $\langle \mathcal{S, A, O, P, R, \gamma}\rangle$. 
Let $\mathcal{S}$ be the global state, $\mathcal{A} = \prod_{i=1}^N \mathcal{A}^i$ the joint action space, and $\mathcal{O} = \prod_{i=1}^N \mathcal{O}^i$ the joint observation space. $\mathcal{O}^i$ partially observes $\mathcal{S}$. $\mathcal{P}$ describes the environment's transition probabilities, $\mathcal{R}$ the reward function for each agent and $\gamma$ the discount parameter.

\subsubsection{State, transition probabilities and actions}
The state of the environment $X \in \mathcal{S}$ and its transition probabilities $\mathcal{P}$ are unknown to the agent. They are simulated by the environment dynamics described in Section~\ref{sec:environment}.
Each agent $i$'s action $a^i_{t} \in \mathcal{A}^i$ is a binary decision to control the AC status. If the remaining lockout time $l^i_{t}$ is above zero, the \textsc{on} action will be ignored by the AC. In practice, a backup controller within the AC would prevent the \textsc{on} decision from being implemented.

\subsubsection{Observations and communications}
\label{sec:comm_POMDP}

By default, agent $i$ receives observation  $o^i_t = \{T^i_{h,t}, T^i_{m,t}, T^i_T, \omega^i_t, l^i_{t}, s_t/N, P_t/N\}$ at time step $t$, where $T^i_{h,t}$, $T^i_{m,t}$ and $T^i_T$ are the indoor air, mass, and target temperatures, $\omega^i_t$ is the \textsc{on} or \textsc{off} status of the AC, $l^i_{t}$ is its remaining lockout time,  $s_t/N$ is the per-agent regulation signal and $P_t/N$ is the per-agent total consumption of the aggregation.

Each agent $i$ communicates with its $N_c$ neighbours. The messages' sizes are not hard limited but should be small, and their contents are not constrained. We define the set of all of agent $i$'s $N_c$ neighbours as $M^i$. By default, we organize the agents in a 1-dimensional structure: $
    M^i = \{i - \lfloor N_c/2 \rfloor,  i - \lfloor N_c/2 \rfloor +1, \ldots, i,  \ldots, i + \lfloor N_c/2\rfloor  -1,    i + \lfloor N_c/2\rfloor \} \backslash \{i \}
$.

\subsubsection{Reward}
For each agent $i$, reward $r_t^i$ is computed as the weighted sum of the penalties due to its air temperature difference with the target, which is unique to the agent, and to signal tracking, which is common across all agents. This scenario is therefore cooperative with individual constraints. We normalize the reward with $\alpha_{\mathrm{temp}} = 1$ and $\alpha_{\mathrm{sig}} = 3 \times 10^{-7}$: a 0.5 $\degree$C error is penalized as much as a 912 W per-agent error (each agent consumes 6000 W).
\begin{equation*}
    r^i_{t} = - \left(\alpha_{\mathrm{temp}}\left( T^i_{h,t} - T^i_{T,t} \right)^2 + \alpha_{\mathrm{sig}} \left(\dfrac{ P_{t} - s_t}{N}\right)^2 \right)
\end{equation*}

%% file: Contents/4_Methods.tex
\section{Classical and learning-based algorithms}
\label{sec:methods}

\subsection{Classical baselines}
\label{sec:classical_baselines}

To the best of our knowledge, there is no classical baseline that performs well under all the constraints enumerated in Section \ref{sec:introduction}. However, simple algorithms can optimize selected objectives, and we use them as baselines for the results of the MARL agent.

\subsubsection{Bang-bang controller}
\label{sec:bangbang}
The bang-bang controller (BBC) turns the AC \textsc{on} when the air temperature $T^i_{h,t}$ is higher than the target $T^i_T$, and \textsc{off} when it is lower. This is a decentralized algorithm, which does not consider demand response but near-optimally controls the temperature.
When the lockout duration $l^i_{\mathrm{max}}$ is 0, the BBC optimally controls the temperature, but does not account for the signal. As the base signal $s_{0,t}$ is computed to allow optimal temperature control, BBC's signal tracking error is mainly due to the high-frequency variations of the signal.

\subsubsection{Greedy myopic}
\label{sec:greedymyopic}

The greedy controller is a centralized algorithm that solves a knapsack problem~\cite{Dantzig_1957} where the size of the collection is the regulation signal, the weight of each AC is its consumption $P^i_{h,t}$, and its value is the temperature difference $T^i_{h,t} - T^i_T$. At each time step, ACs are chosen based on a value priority computed by $(T^i_{h,t} - T^i_T)/P^i_{h,t}$, until the aggregation's consumption $P_t$ is higher than the regulation signal $s_t$.
As it does not plan for the future, the greedy myopic approach quickly runs out of available ACs as most of them are in lockout. However, with a 0-lockout duration $l^i_{\mathrm{max}}$, it is optimal to track the signal $s_t$, and controls the temperature in second priority. We implement the greedy myopic approach as it is better adapted to these settings than the OO approach described in Section \ref{sec:relatedworks}. Indeed, OO only uses past state information and must be implemented in a strictly online fashion. Both frameworks are myopic, and struggle similarly with the lockout constraint.

\subsubsection{Model predictive control}
\label{sec:MPC}
Model predictive control, or MPC, is in its nominal form a centralized algorithm modeling the environment and identifying the actions which will lead to the highest sum of rewards over a time horizon of $H$ time steps.
As the signal is stochastic, MPC assumes a constant future signal over horizon $H$, and optimally solves the trajectory with lockout. However, because it is a large-scale combinatorial optimization problem, it scales poorly with the number of agents $N$ and with a horizon $H$. In the best case the complexity is polynomial, but it is exponential in the worst case. As a result, we were not able to run the MPC for more than 10 agents for $H=60$s, and had to increase the time step between each action to 12 seconds. More details are provided in Appendix~\ref{sec:app_MPC}. 

\subsection{Learning-based methods}
We deploy two algorithms using deep reinforcement learning, namely MA-DQN and MA-PPO, both using the CT-DE paradigm. While MA-DQN only uses hand-engineered communications, MA-PPO was implemented with two communications paradigms: hand-engineered and learned. Details about the architectures and hyperparameters are provided in~Appendix \ref{sec:app_learning_based}.

\subsubsection{Centralized Training, Decentralized Execution}
The CT-DE paradigm \cite{Kraemer_Banerjee_2016} assumes that information is shared during the training of the agents, while they execute actions only based on their decentralized observations. This reduces the non-stationarity of the environment \cite{Gronauer_Diepold_2021} and stabilizes the training. In our case, all agents are homogeneous, which allows the use of parameter sharing \cite{Gupta_Egorov_Kochenderfer_2017}.
As such, all ACs are controlled by identical instances of the same policy trained from the shared experience of all agents.

\subsubsection{MA-DQN}
Multi-agent Deep Q-Network (MA-DQN) is the CT-DE adaptation of DQN \cite{Mnih_DQN}, an off-policy algorithm made for discrete action spaces.  
A DQN agent mainly consists of a $Q$-network predicting the $Q$-value of action-observation pairs $(a^i_t, \Tilde{o}^i_t)$ for every possible $a^i_t$. During training, at time step $t$, the transition $\Theta^i_t = \{\Tilde{o}^i_t, a^i_t, r^i_t, \Tilde{o}^i_{t+1}\}$ of every agent is recorded in a common replay buffer. This replay buffer is sampled to train the $Q$-network to predict $Q(a^i_t, \Tilde{o}^i_t)$ supervised with target $T(a^i_t, \Tilde{o}^i_t)$ according to Bellman's optimality equation:
\begin{equation*}
T(a^i_t, \Tilde{o}^i_t) = r^i_t + \gamma \max_{a} Q(a,\Tilde{o}^i_{t+1}).
\end{equation*}
Actions are selected as $a^i_t$ with maximal predicted $Q$-value given an input $\Tilde{o}^i_t$. $\epsilon$-greedy exploration is added during training. 

\subsubsection{MA-PPO}
Multi-agent Proximal Policy Optimization (MA-PPO) \cite{yu2021surprising} is the CT-DE adaptation of clipped PPO \cite{Schulman_Wolski_Dhariwal_Radford_Klimov_2017}, an on-policy, policy-gradient algorithm. 
The agent jointly learns a policy $\pi_\theta(a^i_t|\Tilde{o}^i_t)$, also called an actor, and a value function $V_\phi(\Tilde{o}^i_t)$, also called a critic.
At each epoch, the policy is fixed and the transitions $\Theta^i_t = \{\Tilde{o}^i_t, a^i_t, \pi_{\theta_t}(a^i_t|\Tilde{o}^i_t), r^i_t\}$ for all agents are recorded together for one or several episodes of length $H$. For each $\Theta^i_t$, a return $G^i_t = \sum_{\tau = 0}^{H-t} \gamma^{\tau} r^i_{t+\tau}$ is computed based on future experience. Then, the new policy parameters $\theta_{t+1}$ are trained over the stored memory to optimize the clipped PPO objective $\mathcal{L}(\Tilde{o}^i_t, a^i_t, \theta_{t+1}, \theta_t)$, maximizing the advantage $A^{\pi_{\theta_t}}(\Tilde{o}^i_t, a^i_t) = G^i_t - V_{\phi}(\Tilde{o}^i_t)$ under the constraint of proximity around the previous policy.
The critic parameters $\phi$ are then trained so that $V_{\phi}(\Tilde{o}^i_t)$ predicts the return $G^i_t$. The memory is erased and a new epoch starts.

Exploration is handled by the inherent stochasticity of the policy.
In the CT-DE setting, $V_\phi$, which is only used during training, is given additional information about the states of other agents.

\subsubsection{Communications}
\paragraph{Hand-engineered communications}
For MA-DQN and the hand-engineered MA-PPO, the messages are designed based on the state of each agent, effectively providing a wider observability of the general state. Agent $j$'s message $m_{j,t}$ contains the current difference between its air and target temperatures $T^j_{h,t} - T^j_T$, its remaining lockout time $l^j_t$, and its current status $\omega^j_t$.
The messages $\{m^i_{j,t}\}_{\forall j \in M_i}$ from agents $j \in M^i$ are concatenated with the observations $o^i_t$ to create the input $\Tilde{o}^i_t$ of the neural networks. Message $m^i_{j,t}$ from agent $j$ to agent $i$ is at a fixed place in the $\Tilde{o}^i_t$ vector based on its relative position $i-j$. MA-PPO with hand-engineered communication will be referred to as MA-PPO-HE.

\paragraph{Targeted Multi-Agent Communication} To allow agents to learn to communicate, we implement TarMAC \cite{Das_2019} in MA-PPO. TarMAC is an attention-based targeted communication algorithm where each agent outputs a key, a message and a query. The key is sent along with the message to the other agents, which then multiply it with their query to compute the attention they give to the message. All messages are then aggregated using the attention as a weight. The three modules -- key, message, query -- are trained. TarMAC allows more flexibility to the agents: it does not restrict the contents of the communication, and it allows agents to communicate with a different number of houses than they were communicating with during training. More details are available in Appendix~\ref{sec:App_TarMAC}. We refer to this version as TarMAC-PPO.

\paragraph{No communication}
It is also possible to train agents without communication. In this case, it only observes $o^i_t$. This agent is referred to as MA-PPO-NC.

\subsubsection{Agent training}
The learning agents were trained on environments with $N_{\mathrm{tr}} = \{10, 20, 50\}$ houses and communicating with $N_{\mathrm{c}_{\mathrm{tr}}}  = \{9, 19, 49\} $ other agents. We trained every agent on 16 different seeds: 4 for environment and 4 for network initialization. They were trained on 3286800 time steps, equivalent to 152 days, divided in 200 episodes. Each episode is initialized with each house having a temperature higher than the target, sampled from the absolute value of a 0-mean Gaussian distribution with $\sigma=5\degree$C. We tuned the hyperparameters through a grid search, as shown in Appendix~\ref{sec:app_algodetails}. The contribution of this paper is to demonstrate that learning-based methods can lead to high performance on the problem of high frequency regulation. We therefore do not compile statistics over the trained agents; instead, for each situation, we select the two best agents over the seeds based on test return, and report the best score from these two on the benchmark environment.

%% file: Contents/5_Results.tex
\section{Results and analysis}
\label{sec:results}

\subsection{Metrics of performance}
We deploy the agents on a benchmark environment with $N_\mathrm{de}$ houses on trajectories of 43200 steps, i.e., two full days. We evaluate their performance with the per-agent root mean square error (RMSE) between the regulation signal $s_t$ and aggregated power consumption $P_t$. We also measure the temperature RMSEs -- one for all agents, one of the maximal temperature error of the aggregation -- to ensure thermal control. Every house's temperature is initialized differently, so we start computing the RMSE when the temperature is controlled, after 5000 steps. For context, a single AC consumes 6000 W when turned \textsc{on}. 
Due to the MPC's computing time, its performance is evaluated differently, as explained in Appendix~\ref{sec:app_MPC}.
Unless mentioned otherwise, the results are the mean and standard deviation over 10 environmental seeds.

\subsection{Performance of agents}
Table \ref{tab:performance} shows the performance of different agents in environments with and without lockout with $N_\mathrm{de}$ of 10, 50, 250 and 1000 houses.

\begin{table*}
    \centering
    \caption{Performance of the different agents, computed over 10 environment seeds.}
    \begin{adjustbox}{max width=0.9\linewidth}
    \begin{tabular}{|c|c|c|c|c|c|c|c|c|c|c|c|c|c|}
    \hline
    \multicolumn{2}{|c|}{ } & \multicolumn{3}{|c|}{$N_\mathrm{de} = 10$} & \multicolumn{3}{|c|}{$N_\mathrm{de} = 50$} & \multicolumn{3}{|c|}{$N_\mathrm{de} = 250$} & \multicolumn{3}{|c|}{$N_\mathrm{de} = 1000$} \\
    \hline
    \multicolumn{2}{|c|}{Per-agent} & Signal& T. & Max T. & Signal & T.  & Max T.  & Signal& T. & Max T. & Signal & T.  & Max T.\\
    \multicolumn{2}{|c|}{RMSE} & (W) & (\degree C) & (\degree C) & (W) & (\degree C) & (\degree C) & (W) & (\degree C) & (\degree C) & (W) & (\degree C) & (\degree C) \\
    \hline
    \multirow{2}{*}{No l.o} & Greedy& 194 $\pm$ 1 & 0.04    & 0.06    & 70 $\pm$ 1 & 0.03    & 0.05  & 63 $\pm$ 1 & 0.03 & 0.052 & 63 $\pm$ 1 & 0.03 & 0.05 \\
    & BBC  & 806 $\pm$ 147 & 0.02    & 0.03    & 392 $\pm$ 50 & 0.02    & 0.04  & 310 $\pm$ 11 & 0.02 & 0.03 & 272 $\pm$ 12 & 0.02 & 0.03   \\
    \hline
    \multirow{7}{*}{40s l.o} & Greedy& 2668 $\pm$ 14 & 0.87    & 0.93    & 3166 $\pm$ 12 & 1.09    & 1.15   & 3313 $\pm$ 12 & 1.16 & 1.22 & 3369 $\pm$ 15 & 1.18 & 1.24  \\
    &BBC & 830 $\pm$ 207 & 0.05    & 0.09    & 426 $\pm$ 63 & 0.05    & 0.10  & 318 $\pm$ 7 & 0.05 & 0.10 & 296 $\pm$ 4 & 0.05 & 0.10   \\
    &MPC & 344 $\pm$ 96 & 0.07 & 0.12 & - & - & - & - & - & - & - & - & - \\
    &MA-DQN & 541 $\pm$ 86 & 0.05    & 0.09    & 321 $\pm$ 24 & 0.05    & 0.10 & 246 $\pm$ 8 & 0.05 & 0.11 & 234 $\pm$ 4 & 0.05 & 0.12  \\
    &MA-PPO-HE & 253 $\pm$ 1 & 0.04    & 0.08    & 161 $\pm$ 8 & 0.04    & 0.08 & 127 $\pm$ 2 & 0.04 & 0.11 & 122 $\pm$ 3 & 0.05 & 0.13 \\
    &TarMAC-PPO & \textbf{247 $\pm$ 3} &\textbf{ 0.04} &\textbf{ 0.07} & \textbf{158 $\pm$ 2} & \textbf{0.04} & \textbf{0.09} & \textbf{115 $\pm$ 1} & \textbf{0.05} & \textbf{0.13} & \textbf{101 $\pm$ 2} & \textbf{0.05} & \textbf{0.14}\\
    & MA-PPO-NC & 434 $\pm$ 2 & 0.06 & 0.08 & 215 $\pm$ 1 & 0.06 & 0.14 & 132 $\pm$ 1 & 0.06 & 0.16 & 107 $\pm$ 1 & 0.06 & 0.17 \\
    \hline
    \end{tabular}
    \end{adjustbox}
    \label{tab:performance}
\end{table*}

The per-agent signal RMSE generally goes down when $N_\mathrm{de}$ increases. This is due to the lower relative discretization error, but also because, with more agents, errors have more chances to cancel each other, as explained in Appendix~\ref{sec:app_proof_smallererror}. 
As expected, BBC controls the temperature well, but does not track the signal. Without lockout, the greedy myopic shows near-optimal signal tracking, where errors are due to discretization. It also maintains good control of the temperature. With lockout, however, it fails, as it runs out of available agents. The MPC gives good results for 10 agents, but its performance is limited by the lower control frequency of 12 seconds. It could not be run on $N_\mathrm{de}=50$ for computing time reasons. DQN controls the temperature well but is only slightly better than BBC on the signal. Both PPO agents show significantly better performance, and TarMAC-PPO outperforms MA-PPO-HE at high $N_{\mathrm{de}}$.  
The results without communication will be discussed in Section~\ref{sec:results_comm}.

\begin{figure*}
    \centering
    \includegraphics[width=0.87\textwidth]{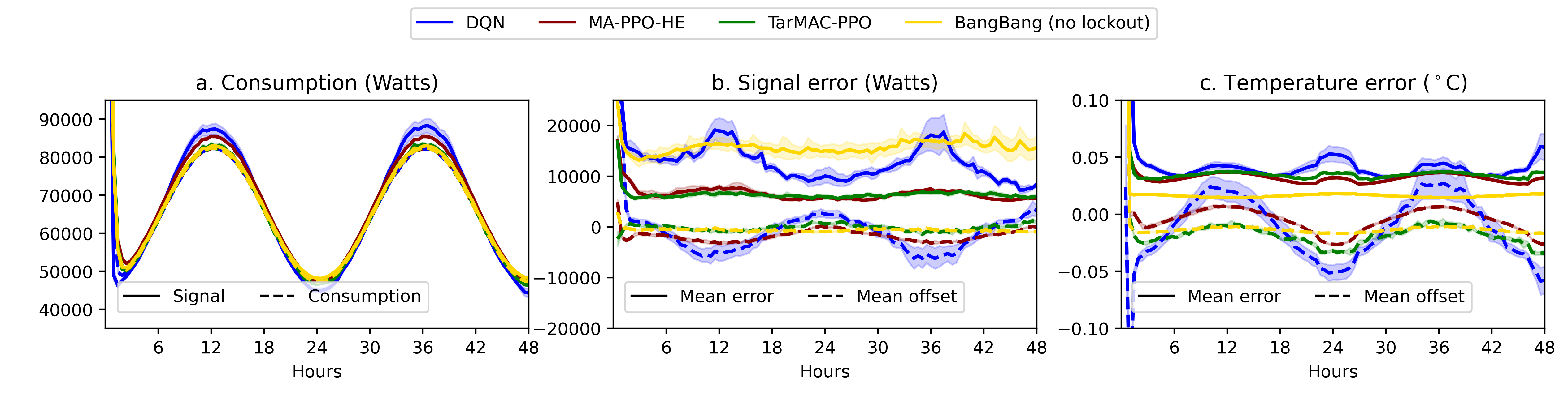}
    \caption{MA-PPO-HE and TarMAC-PPO outperform DQN and BBC for signal and temperature over 2 days with $N_\mathrm{de} = 50$ agents.}
    \label{fig:performance_50}
\end{figure*}

Figure~\ref{fig:performance_50} shows the behaviour of each agent over two days for 50 houses. Every point on the curves is averaged over 10 minutes. The mean offset captures the error's bias by averaging the differences such that positives and negatives cancel each other, while the mean error is the mean of the absolute differences. 
The signal and consumption curves start very high due to the initial situation, and then follow the sinusoidal pattern of the outdoor temperature. Without lockout, the BBC shows low temperature and signal offsets, with a significant signal error, as it does not track high-frequency variations of the signal. With the lockout, it under-consumes as explained in Section~\ref{sec:bangbang}, leading to a positive temperature offset, and the base signal rises to compensate. As the signal variation amplitude is high, this does not strongly affect the error.

The DQN agent has a smaller signal offset and error, especially at night when the amplitude of the signal variations is lower. During the day, the signal error is still significant. Both MA-PPO agents, on the other hand, have a near-0 offset in signal and temperature. Their signal error is also significantly lower than the others, because they are able to track the high-frequency variations. 

\subsection{Scalability with number of agents}
\begin{figure*}
    \centering
    \includegraphics[width=0.87\textwidth]{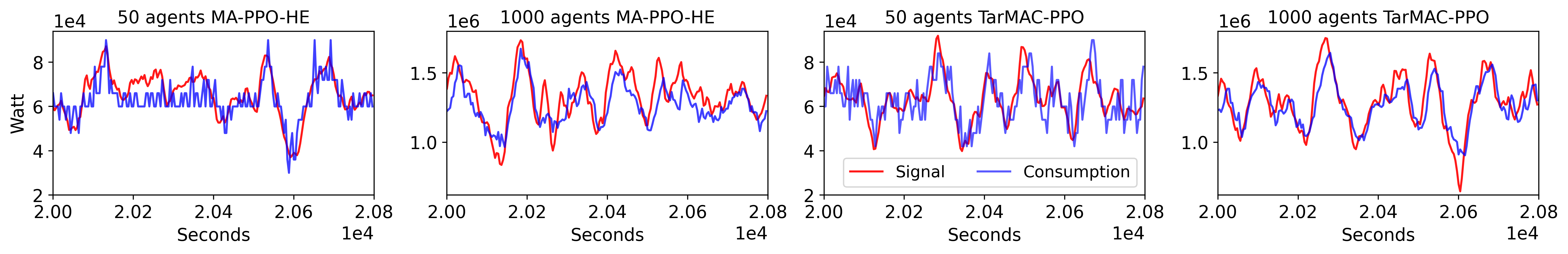}
    \caption{Both MA-PPO policies scale seamlessly in the number of agents: signal and consumption on 800s for $N_\mathrm{de}= 50$ and $1000$.}
    \label{fig:performance_shortterm}
\end{figure*}

\begin{figure}[ht]
    \centering
    \includegraphics[width=0.43\textwidth]{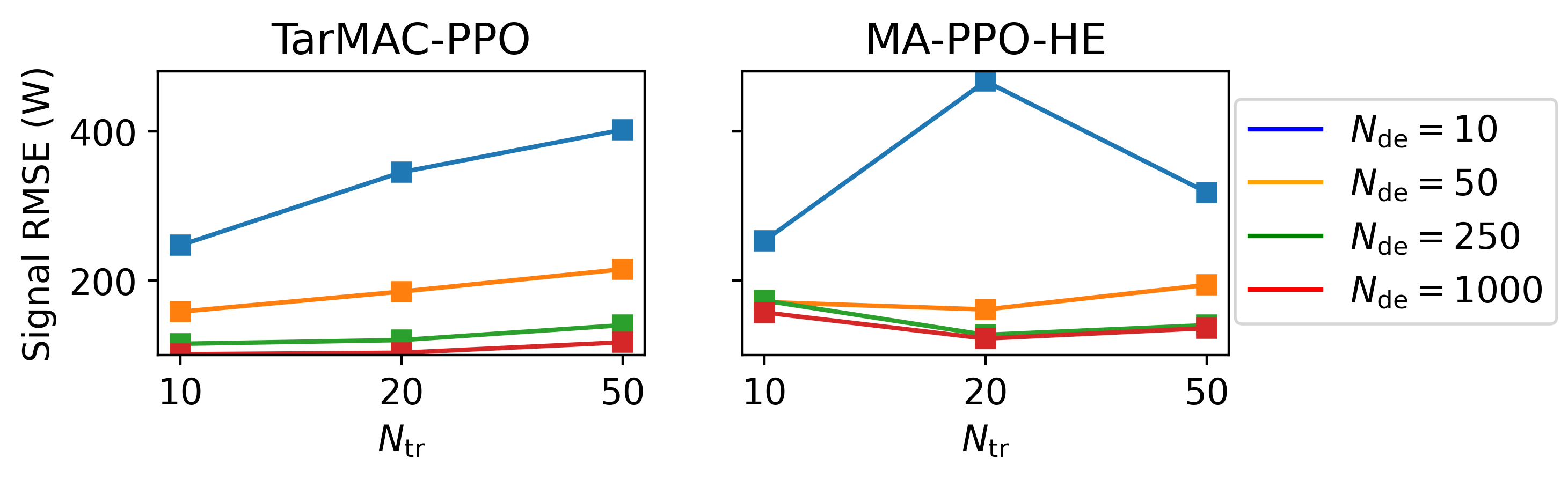}
    \caption{Training with more agents $N_\mathrm{tr}$ does not lead to better performance, even when deployed on large $N_\mathrm{de}$.}
    \label{fig:Ntr_Nde}
    \vspace{-0.4cm}
\end{figure}

As shown in Table~\ref{tab:performance}, the PPO agents, and TarMAC-PPO especially, scale gracefully with the number of agents.
Figure~\ref{fig:performance_shortterm} shows the consumption and signal over 800 seconds for agents deployed over $N_d = 50$ and 1000 over $800$ seconds. For $N_d = 50$, the agents do not perfectly match the signal. However, the same agent does better on 1000 houses. Indeed, as the environment is homogeneous, the local strategy scales smoothly by averaging out errors.  The best performing agents for TarMAC-PPO were trained on environments with $N_\mathrm{tr} =$ 10 houses. With MA-PPO-HE, it is often the agents trained on $N_\mathrm{tr} =$ 20 that had the best results. Training with $N_\mathrm{tr} =$ 50 probably makes the credit assignment harder as shown in Figure~\ref{fig:Ntr_Nde}.

\subsection{PPO agents' dynamics}

\begin{figure*}[ht]
     \centering
     \begin{subfigure}[b]{0.49\textwidth}
         \centering
         \includegraphics[width=\textwidth]{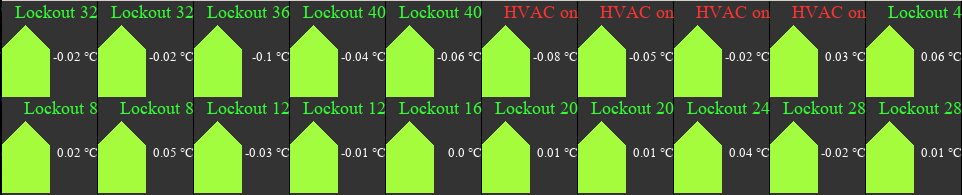}
         %\caption{This agent trained on 20 houses with $N_c = 9$ developed a "3-house" pattern.}
     \end{subfigure}
     \hfill
     \begin{subfigure}[b]{0.49\textwidth}
         \centering
         \includegraphics[width=\textwidth]{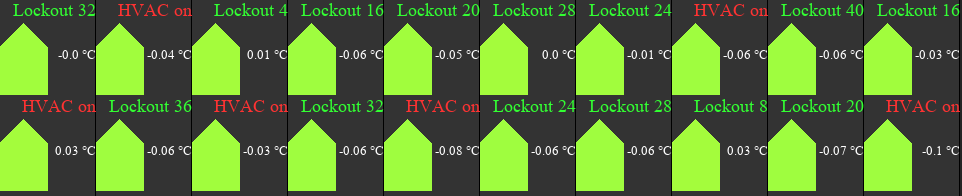}
         %\caption{This agent trained on 20 houses with $N_c = 19$ instead developed a "20-house" pattern.}
     \end{subfigure}
     \hfill 
     \begin{subfigure}[b]{0.49\textwidth}
         \centering
         \includegraphics[width=\textwidth]{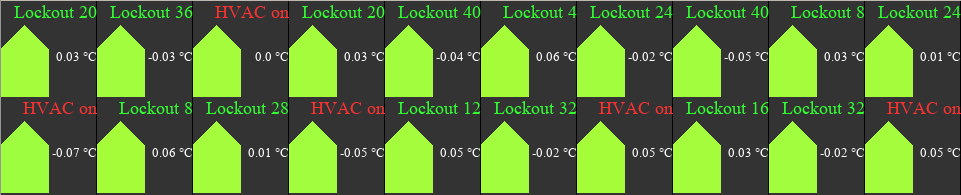}
         \caption{MA-PPO-HE}
     \end{subfigure}
     \hfill
     \begin{subfigure}[b]{0.49\textwidth}
         \centering
         \includegraphics[width=\textwidth]{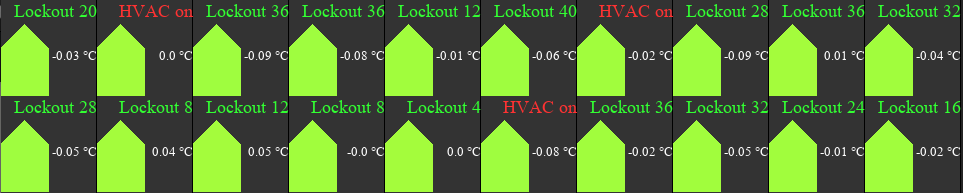}
         \caption{TarMAC-PPO}
     \end{subfigure}     
        \caption{State of 20 houses controlled with two different PPO agents. The number on the top right is the remaining lockout time. (Left) Two different agents of MA-PPO-HE with $N_{\mathrm{c}_\mathrm{de}} = 19$ show  a ``20-house'' (up) and a ``3-house'' (down) pattern. (Right) Two different TarMAC-PPO agents show no such pattern.}  
        \label{fig:viz_20}
\end{figure*}

As visualized in Figure~\ref{fig:viz_20}, both MA-PPO-HE and TarMAC-PPO policies keep the ACs in lockout or \textsc{on}, and never \textsc{off}. This is optimal for temperature control: an agent needing to be \textsc{off} to warm up after lockout, would not have had the time to warm up during the lockout and was thus \textsc{on} for too long beforehand. The agents turn \textsc{on} as soon as they can, but control when they turn \textsc{off} based on the context and the messages of other agents.

A fascinating feature of the learned policies is the cyclic behaviour used by MA-PPO-HE agents for coordination. As shown in Figure~\ref{fig:viz_20}, the ACs turn \textsc{on} one after the other based on their positions in the aggregation, with a repetitive pattern. This happened for each MA-PPO-HE agent we trained, although the pattern period or moving direction was different. These patterns enable agent coordination thanks to the stable message structure, i.e., the fixed relative position of agent $j$'s message to agent $i$ in the $\Tilde{o}^i_t$ vector. 
The TarMAC-PPO agents, on the other hand, do not follow a pattern in their collective behavior. Indeed, aggregated messages do not contain information about the structure of the neighbours. The coordination is done through flexible message contents.

\subsection{Communications}
\label{sec:results_comm}
\begin{figure}[ht]
    \centering
    \includegraphics[width=0.49\textwidth]{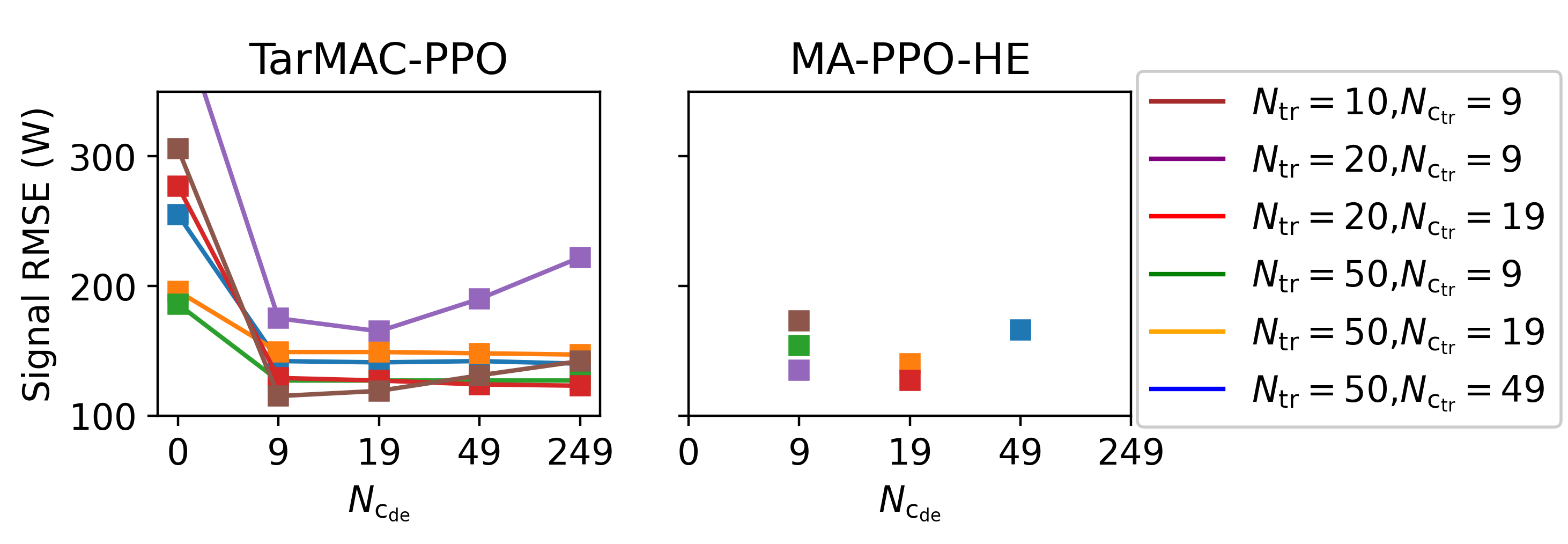}
    \caption{TarMAC-PPO's performance does not increase after $N_{\mathrm{c}_{\mathrm{de}}} = 9$, while MA-PPO-HE is better with $N_{\mathrm{c}_{\mathrm{de}}} = 19$, for $N_\mathrm{de} = 250$ agents.}
    \label{fig:Nctr_Ncde}
\end{figure}

\begin{figure}[ht]
    \centering
    \includegraphics[width=0.3\textwidth]{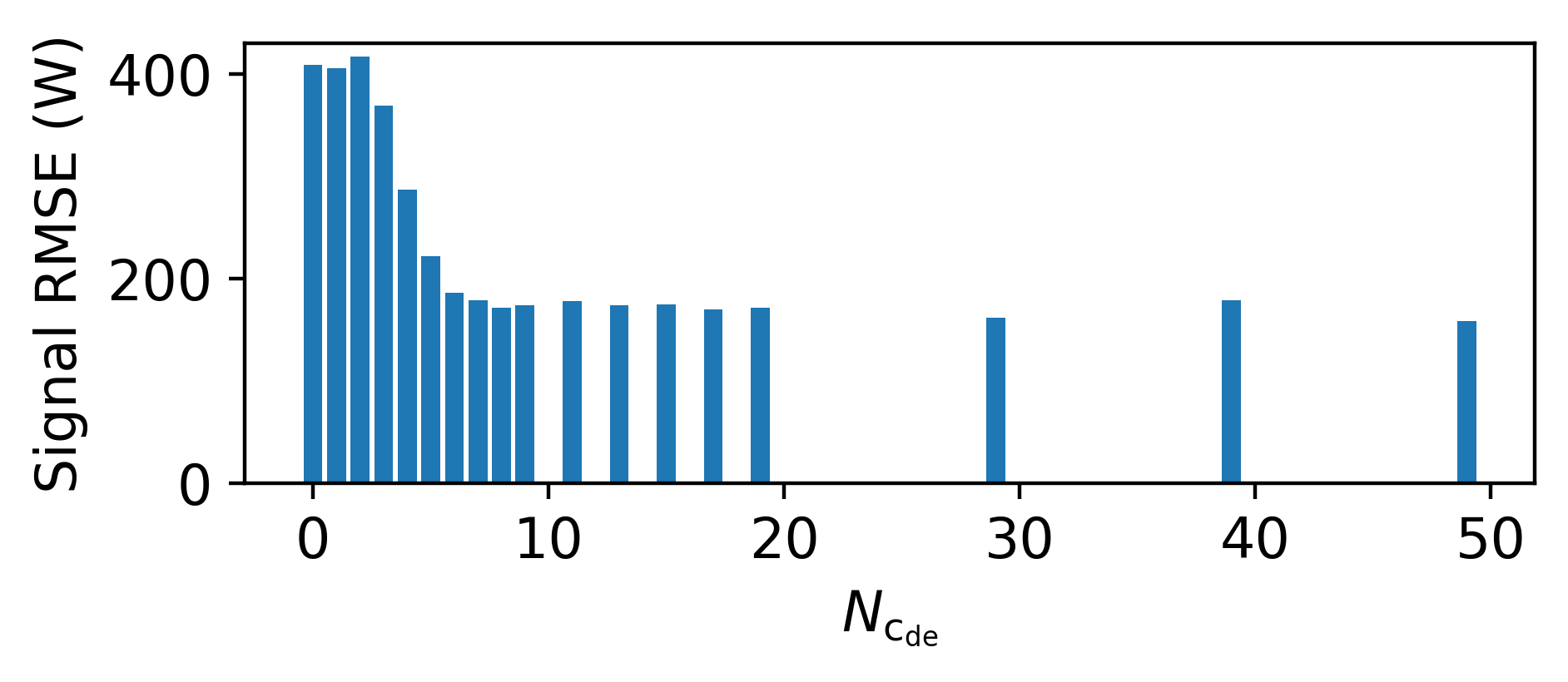}
    \caption{A TarMAC-PPO agent performs well as long as it communicates with $N_{\mathrm{c}_{\mathrm{de}}} = 7$ agents or more, on $N_\mathrm{de} = 50$.}
    \label{fig:LowNcde}
\end{figure}

The agents need communications to coordinate and get the best results. Intuitively, the more agents to communicate with, the better the performance because the observability of the environment is improved. In practice, this is not always the case, as shown in Figure~\ref{fig:Nctr_Ncde}.
For TarMAC-PPO, communicating with 9 neighbours often leads to the best performance. Higher values of $N_{\mathrm{c}_{\mathrm{de}}} $ can lead to a reduction of the weight of important messages in the aggregation. For MA-PPO-HE, communicating with 19 agents yields better results than with 49. Indeed, in MA-PPO-HE, the agents must have $N_{\mathrm{c}_{\mathrm{de}}}  = N_{\mathrm{c}_{\mathrm{tr}}} $. During training, communicating with more agents increases the credit assignment difficulty as it increases the input size with non-controllable elements. 
It is also clear in Figure~Figure~\ref{fig:Nctr_Ncde} that agents trained to communicate do not cope well when not communicating. Figure~\ref{fig:LowNcde} shows the performance of a TarMAC agent trained with $N_\mathrm{tr} = 10$ and $N_{\mathrm{c}_{\mathrm{tr}}} =9$ on an environment with $N_\mathrm{de} = 50$ agents, when changing the number $N_{\mathrm{c}_{\mathrm{de}}} $ of neighbours it can communicate with. The performance is bad at low communication but stabilizes around 7 or 8 agents.

It is, however, possible to train an agent without communication to do better than Bang-Bang control, as shown by the performance of MA-PPO-NC in Table~\ref{tab:performance}. Without coordinating with the others, an agent can learn to act well on average to minimize the signal error. When there are only a few agents, as when $N_\mathrm{de} = 10$ or 50, this does not perform very well. However, the performance gap decreases when $N_\mathrm{de}$ increases: a good average policy will do well when applied on many agents. Another way to see this is that, with large $N_\mathrm{de}$, each agent's importance becomes negligible in the final result. As such, the group can be seen as a single average agent, and the problem cam be posed as a mean field game \cite{Yang_Luo_Li_Zhou_Zhang_Wang_2018, Subramanian_Seraj_Mahajan_2018}. 

Interestingly, MA-PPO-HE at high $N_\mathrm{de}$ does better with communication defects. This may be because the MA-PPO-HE coordination leads to locally biased policies, which do not benefit from the averaging effect reducing the relative error when $N_\mathrm{de}$ increases.

\subsection{Robustness}
All the results presented were produced under certain assumptions, such as homogeneous houses and ACs, consistent outdoor temperature and signal profiles, and faultless communication. 
If such agents were to be deployed in the real world, they would be confronted with situations where these conditions are not satisfied. In this section, we evaluate the robustness of our trained agents to different disturbances in the deployment conditions. 

\subsubsection{Faulty communications}
As previously demonstrated, communications are key for good performance of the agents. In this robustness test, we simulate defective communications. At every time step, each message $m^i_j$ is defective with a probability $p_d$. In the case of TarMAC-PPO, this leads to the message not being received. For MA-PPO-HE, every element of the message is set to 0.
We tested the best agents for $N_\mathrm{de} = 10$, 50, 250 and 1000 houses with $p_d = 0.1$ and 0.5, as seen in Table~\ref{tab:robustness_comm}.
MA-PPO-HE agents' coordination is based on their stable communication structure. As a result, it copes badly with defective communications. Interestingly, when $N_\mathrm{de}$ is higher, the impact decreases, even leading to better performance at $N_\mathrm{de} = 1000$. This may be due to the fact that the resulting policies cannot coordinate locally and are less locally biased.
The TarMAC-PPO handles perfectly temporary defects in communication as its messages are aggregated. This is the case even with $p_d = 0.5$ and when the agent communicates with $N_{\mathrm{c}_{\mathrm{tr}}}  = 9$ neighbours only. 

\begin{table*}
    \centering
    \caption{Performance under faulty communication (5 seeds)}
    \begin{adjustbox}{max width=0.9\linewidth}
    \begin{tabular}{|c|c|c|c|c|c|c|c|c|c|c|c|c|c|}
    \hline
    \multicolumn{2}{|c|}{ } & \multicolumn{3}{|c|}{$N_\mathrm{de} = 10$} & \multicolumn{3}{|c|}{$N_\mathrm{de} = 50$} & \multicolumn{3}{|c|}{$N_\mathrm{de} = 250$} & \multicolumn{3}{|c|}{$N_\mathrm{de} = 1000$} \\
    \hline
    \multicolumn{2}{|c|}{Per-agent} & Signal& T. & Max T. & Signal & T.  & Max T.  & Signal& T. & Max T. & Signal & T.  & Max T.\\
    \multicolumn{2}{|c|}{RMSE} & (W) & (\degree C) & (\degree C) & (W) & (\degree C) & (\degree C) & (W) & (\degree C) & (\degree C) & (W) & (\degree C) & (\degree C) \\
    \hline
    \multirow{3}{*}{MA-PPO-HE} & $p_d = 0$ & 253 $\pm$ 1 & 0.04  & 0.08   & 161 $\pm$ 8 & 0.04    & 0.08  & 127 $\pm$ 2 & 0.04 & 0.11 & 122 $\pm$ 3 & 0.05 & 0.13 \\
    & $p_d = 0.1$ & 504 $\pm$ 2 & 0.07   & 0.14   & 207 $\pm$ 1 & 0.04  & 0.11  & 138 $\pm$ 2 & 0.05 & 0.13 & 118 $\pm$ 1 & 0.05 & 0.14 \\
    & $p_d = 0.5$ & 597 $\pm$ 2 & 0.10  & 0.19   & 274 $\pm$ 1 & 0.06 & 0.15 & 148 $\pm$ 1 & 0.06 & 0.151 & 115 $\pm$ 2 & 0.06 & 0.17 \\
    \hline
    \multirow{3}{*}{TarMAC-PPO} & $p_d = 0$  & 247 $\pm$ 3 & 0.04  & 0.07   & 158 $\pm$ 2 & 0.04 & 0.09  & 115 $\pm$ 1 & 0.05 & 0.13 & 101 $\pm$ 2 & 0.05 & 0.14 \\
    & $p_d = 0.1$  & 246 $\pm$ 2 & 0.04  & 0.07 & 158 $\pm$ 2 & 0.04  & 0.09  & 115 $\pm$ 2 & 0.05 & 0.12 & 101 $\pm$ 1 & 0.05 & 0.14 \\
    & $p_d = 0.5$  & 248 $\pm$ 2 & 0.04  & 0.07 & 159 $\pm$ 3 & 0.04 & 0.09  & 115 $\pm$ 2 & 0.05 & 0.13 & 101 $\pm$ 1 & 0.05 & 0.14 \\
    \hline
    \end{tabular}
    \end{adjustbox}
    \label{tab:robustness_comm}
\end{table*}

\subsubsection{Heterogeneous houses and ACs}
In reality, different houses have different thermal characteristics. The ACs also do not always have the same rated power or lockout duration. 
We deployed the best trained MA-PPO-HE and TarMAC-PPO agents for 50-house environments that do not comply with these assumptions, to evaluate their robustness to separate disturbances. We also trained new agents on environments with these conditions, to allow the agents to learn to cope with heterogeneity. The relevant characteristics were observed by both agents as part of $o^i_t$, and of the messages $m^i_j$ in MA-PPO-HE. These agents are referred to with the -T suffix. The thermal characteristics heterogeneity was simulated by adding a Gaussian noise to each element of $\theta^i_h$ for each house, with a standard deviation of 50\% of the original value (the final values cannot be negative).
For the ACs cooling capacities $K_a^i$, a value between 10, 12.5, 15, 17.5 and 20 kW was uniformly selected for each house.
Finally, heterogeneity in the lockout duration $l_\mathrm{max}$ was tested by sampling uniformly between 32, 36, 40, 44 and 48 seconds. 

The results are shown in Table~\ref{tab:Hetero_housesACs}. TarMAC-PPO is much more robust to heterogeneity in agents than MA-PPO-HE. This is because in MA-PPO-HE the coordination scheme is based on the stable dynamics of the agent's neighbours, especially with the lockout duration. TarMAC-PPO is instead more flexible with respect to different dynamics. For both agents, it is possible to reduce the effect of heterogeneity by training the agents on such environments and allowing them to observe the characteristics. This is different for heterogenity on the lockout duration, where TarMAC-PPO did not seem able to train satisfactorily on such conditions.
An interesting observation is that the best TarMAC-PPO results were obtained when communicating with $N_{\mathrm{c}_{\mathrm{tr}}}  = 49$ agents. With heterogeneous agents, more neighbours are needed for a representative input. 

\begin{table}
\centering
\caption{Performance under house and AC heterogeneity}
\begin{adjustbox}{max width=0.8\linewidth}
    \begin{tabular}{|c|c|c|c|c|}
        \hline
        & \multicolumn{2}{|c|}{MA-PPO-HE} & \multicolumn{2}{|c|}{MA-PPO-HE-T} \\
        \hline
   Per-agent & Signal & Max T.    & Signal & Max T.    \\
    RMSE & (W) & ($\degree$C)   & (W) & ($\degree$C)   \\
    \hline
    Homogeneous      & 161 $\pm$ 8 & 0.08 & - & - \\
    House thermal & 285 $\pm$ 8 & 0.17 & 222 $\pm$ 7 & 0.11 \\
    AC cooling  & 292 $\pm$ 3 & 0.15 & 181 $\pm$ 3 & 0.14  \\
    Lockout duration & 324 $\pm$ 9 & 0.15 & 246 $\pm$ 4 & 0.09  \\
    \hline
    &\multicolumn{2}{|c|}{TarMAC-PPO} & \multicolumn{2}{|c|}{TarMAC-PPO-T} \\
    \hline
    Homogeneous &  158 $\pm$ 2 & 0.09 & - & - \\
    House thermal & 184 $\pm$ 2 & 0.12 & 174 $\pm$ 2 & 0.11 \\
    AC cooling  &  187 $\pm$ 2 & 0.16 & 185 $\pm$ 9 & 0.16 \\
    Lockout duration & 192 $\pm$ 3 & 0.09 & 251 $\pm$ 4 &  0.08 \\
    \hline
    \end{tabular}
\end{adjustbox}
\label{tab:Hetero_housesACs}
\end{table}

\subsubsection{Other environments}
\begin{table}
\centering
\caption{Robustness on environment changes (5 seeds)}
\begin{adjustbox}{max width=\linewidth}
    \begin{tabular}{|c|c|c|c|c|}
        \hline
        & \multicolumn{2}{|c|}{MA-PPO-HE} & \multicolumn{2}{|c|}{TarMAC-PPO} \\
        \hline
    Per-agent & Signal & Max T.    & Signal & Max T.    \\
    RMSE & (W) & ($\degree$C)   & (W) & ($\degree$C)   \\
    \hline
    Same as training    & 161 $\pm$ 8 & 0.08 & 158 $\pm$ 2 & 0.09 \\
    Solar gain & 190 $\pm$ 6 & 0.09 & 174 $\pm$ 2 & 0.10 \\
    Outdoor T. + 4$\degree$C  & 203 $\pm$ 4 & 0.11 & 198 $\pm$ 2 & 0.11 \\
    Outdoor T. - 4$\degree$C  & 170 $\pm$ 1 & 0.09 & 184 $\pm$ 2 & 0.12 \\
    Signal average + 30\% & 401 $\pm$ 2 & 0.11 & 302 $\pm$ 2 & 0.14 \\
    Signal average - 30\% & 337 $\pm$ 4 & 0.10 & 317 $\pm$ 1 & 0.11 \\
    Signal noise amplitude + 30\%  & 188 $\pm$ 5 & 0.08 & 179 $\pm$ 3 & 0.09 \\
    Signal noise frequency + 100\% & 200 $\pm$ 4 & 0.08 & 198 $\pm$ 5 & 0.09 \\
    
    \hline
    \end{tabular}
\end{adjustbox}
\label{tab:Robustness_other}
\end{table}
We also tested our agents on environments differing from the training environment, with different outdoor temperature $T_o$, solar gain $Q_s$, too low or high average signal $D_a$, and higher or faster signal variations $\delta_s$.
As can be seen in Table~\ref{tab:Robustness_other}, both agents are quite robust to such changes, with TarMAC-PPO usually leading to better results. When the signal is misbehaved, i.e., it is too low or too high to allow correct control of the temperature, there is a tradeoff between the signal and the temperature objectives. MA-PPO-HE gives higher priority to temperature, leading to higher signal RMSE.
\balance
\subsection{Processing time}
In Table \ref{tab:processing_time1}, we report the processing time for action selection of the baseline and trained agents. The results are shown for 25 times steps (100 seconds of simulation), except for the MPC which simulated 100 seconds with 10-time steps. They were computed on the 12-core, 2.2 GHz Intel i7-8750H CPU of a laptop computer. 
\begin{table}[h!]
    \centering
    \caption{Computation time (s) for action selection, for 100 seconds of simulation. We report the time per-agents for a decentralized system and for the whole system otherwise.}
    \begin{tabular}{|c|c|c|c|c|c|}
    \hline
        Agent & Decentralized & $N_\mathrm{de} =10$ &  $N_\mathrm{de} =1000$ \\
        \hline
        TarMAC-PPO & Yes & 0.002 & 0.001 \\
        MA-PPO-HE & Yes & 0.006 & 0.006 \\
        DQN & Yes & 0.003 & 0.002 \\
        BBC & Yes & 0.00001 & 0.00001 \\
        Greedy myopic & No & 0.1 & 3.7 \\
        MPC - $H = 40$s & No & 92.6680 & - \\
        \hline
    \end{tabular}
    \label{tab:processing_time1}
\end{table}

As the decentralized, learned agents only need a single forward pass in a relatively small neural network, the time for action selection is sufficiently low for control when using 4-second time steps. Centralized approaches such as greedy myopic scale badly with many agents. 
MPC, already simplified with time steps of 12 seconds instead of 4, and a short horizon of 40 seconds, takes an unacceptable amount of time for more than 10 agents.

%% file: Contents/6_Conclusion.tex
\section{Conclusion}

In this paper, we tackle the problem of high-frequency regulation with demand response by controlling discrete and dynamically constrained residential loads equipped with air conditioners with a decentralized, real-time agent trained by MA-PPO. We test two frameworks for local communication -- fixed hand-engineered messages and learned targeted communication. The policies trained with few agents perform significantly better than baselines, scale seamlessly to large numbers of houses, and are robust to most disturbances. Our results show that MARL can be used successfully to solve some of the complex multi-agent problems induced by the integration of renewable energy in electrical power grids. Future works towards the application of such algorithms on real power systems could include sim2real transfer, integration of more complex flexible loads, as well as power grid safety issues.  

%% file: Contents/Appendix/A_Notation.tex
\flushbottom
\newpage
\section{Notation}
\label{sec:app_notation}

Table \ref{tab:notation} contains the different notations we use in this paper. 

\begin{table*}[ht]
    \centering
    \caption{Notation table}

    \begin{tabular}{|c|c|c|}
    \hline
    \multirow{4}{*}{\centering Number of agents} & $N$ & Number of houses in cluster (general) \\
    & $N_\mathrm{tr}$ & Number of houses in training environment \\
    & $N_\mathrm{de}$ & Number of houses in test environment \\
    & $N_{\mathrm{c}_{\mathrm{tr}}} $ & Number of agents for communication during training\\
    & $N_{\mathrm{c}_{\mathrm{de}}} $ & Number of agents for communication at deployment\\
    \hline
    \multirow{4}{*}{\centering Temperatures} & $T_h$ & Indoor air temperature \\
    & $T_m$ & Indoor mass temperature \\
    & $T_o$ & Outside temperature \\
    & $T_T$ & Target indoor temperature \\
    \hline
    \multirow{9}{*}{\centering Signal and power} & $s_0$ & Base signal \\
    & $\rho$ & Power system operator signal \\
    & $s$ & Regulation signal \\
    & $D_a$ & Average power needed by the ACs \\
    & $D_o$ & Power needed by non flexible loads \\
    & $\delta_s$ & Signal variation \\
    & $\delta_p$ & Perlin noise \\
    & $\beta_p$ & Variation amplitude parameter\\
    & $P$ & Total cluster power consumption \\
    \hline
    \multirow{4}{*}{\centering AC state} & $\omega$ & Status (\textsc{on} or \textsc{off}) \\
    & $l$ & Time left for lockout \\
    & $P_a$ & Power consumption  \\
    & $Q_a$ & Heat removed by the AC \\
    \hline
    \multirow{7}{*}{\centering House thermal model} & $\theta_h$ & House thermal characteristics \\
    & $U_h$ & Outside walls conductance \\
    & $C_m$ & House thermal mass \\
    & $C_h$ & Air thermal mass \\
    & $H_m$ & Mass surface conductance \\
    & $\theta_s$ & House lightning characteristics \\
    & $Q_s$ & Solar gain \\
    \hline
    \multirow{5}{*}{\centering AC model} & $\theta_a$ & AC characteristics \\
    & $K_a$ & Cooling capacity \\
    & $COP_a$ & Coefficient of performance \\
    & $L_a$ & Latent cooling fraction \\
    & $l_{\mathrm{max}}$ & lockout duration \\
    \hline
    \multirow{9}{*}{\centering POMDP} & $a, \mathcal{A}$ & Action, action space \\
    & $o, \mathcal{O}$ & Observation, observation space \\
    & $S, \mathcal{S}$ & State, state space \\
    & $r, \mathcal{R}$ & Reward, reward function \\
    & $\mathcal{P}$ & Transition probabilities \\
    & $M$ & Set of communicating agents \\
    & $m^i_j$ & Message from $j$ to $i$ \\
    & $\Tilde{o}$ & Concatenated observation and messages \\
    & $\gamma$ & Discount factor \\
    & $\alpha_\mathrm{temp}, \alpha_\mathrm{sig}$ & Weights in the reward function \\
    \hline
    \multirow{7}{*}{\centering Algorithms} & $H$ & Horizon \\
    & $\Theta$ & Transition \\
    & $Q(a, o), T(a, o)$ & $Q$-value prediction (with Q or target network) \\
    & $\pi_\theta$ & Policy parameterized by $\theta$ \\
    & $V_\phi$ & Critic parameterized by $\phi$ \\
    & $G$ & Return \\
    & $A^\pi$ & Advantage for policy $\pi$ \\
    \hline
    \end{tabular}
    \label{tab:notation}
\end{table*}

%% file: Contents/Appendix/B_CarbonAccounting.tex
\section{Carbon emissions of the research project}

As a significant amount of electricity has been used to train and run the models for this work, we publish its estimated carbon footprint.

Experiments were conducted using a private infrastructure, which has a carbon efficiency of 0.049 kgCo$_2$eq/kWh. A cumulative of 10895 days, or 261480 hours, of computation was mainly performed on CPU of type Intel Xeon Processor E5-2683 v4 (TDP of 120W). We assume on average a power usage of half the TDP for CPUs.

The total emissions are estimated to be 628 kgCO$_2$eq of which 0\% were directly offset. This is equivalent to 2550 km driven by an average car, or 314 kg of burned coal.

These estimations were conducted using the \href{https://mlco2.github.io/impact/#compute}{MachineLearning}  Impact calculator \cite{Lacoste_Luccioni_Schmidt_Dandres_2019}.

%% file: Contents/Appendix/C_EnvironmentDetails.tex
\section{Environment details}
\label{sec:app_environmentdetails}

\subsection{Detailed house thermal model}
\label{sec:app_house_thermal_model}

The air temperature in each house evolves separately, based on its thermal characteristics $\theta_h$, its current state , the outdoor conditions such as outdoor temperature and solar gain, and the status of the air conditioner in the house. The second-order model is based on Gridlab-D's Residential module user's guide \cite{GridLabD}.

Using Gridlab-D's module, we model an 8$\times$12.5 m, one level rectangular house, with a ceiling height of 2.5 m, 4 1.8-m$^2$, 2-layer, aluminum windows, and 2 2-m$^2$ wooden doors, leading to the following values presented in Table \ref{tab:default_therm_parameters}.

\begin{table}[ht]
    \centering
    \caption{Default house thermal parameters $\theta_h$}
    \begin{tabular}{|c|c|}
    \hline
        $U_h$ & $2.18 \times 10^2$ W$/$K\\
        $C_m$ & $3.45 \times 10^6$ J$/$K\\
        $C_h$ & $9.08 \times 10^5$ J$/$K\\
        $H_m$ & $2.84  \times 10^3$ W$/$K\\
    \hline
    \end{tabular}
    \label{tab:default_therm_parameters}
\end{table}

To model the evolution of the house's air temperature $T_{h,t}$ and its mass temperature $T_{m,t}$, we assume that this temperature is homogeneous and do not consider the heat propagation in the house. We define the following variables:
\begin{align*}
    a =& C_m C_h / H_m \\
    b =& C_m (U_h + H_m)/H_m + C_h \\
    c =& U_h\\
    d =& Q_{a,t} + Q_{s,t} + U_hT_{o,t}\\
    dT_{A0}/dT =& \left(H_mT_{m,t} - (U_h + H_m) T_{h,t}\right. \\ 
    &\left. + U_hT_{o,t} + Q_{h,t} + Q_{s,t}\right)/C_h.
\end{align*}
The following coefficient are then computed: 
\begin{align*}
    r_1 =& (-b + \sqrt{b^2 - 4ac})/2a\\
    r_2 =& (-b - \sqrt{b^2 - 4ac})/2a\\
    A_1 =& (r_2 T_{a,t} - dT_{A0}/dT -  r_2d/c)/(r2-r1)\\
    A_2 =& T_{h,t} - d/c - A_1\\
    A_3 =& (r_1 C_h + U_h + H_m)/H_m\\
    A_4 =& (r_2 C_h + U_h + H_m)/H_m.
\end{align*}
These coefficients are finally applied to the following dynamic equations:
\begin{align*}
    T_{a,t+1} =& A_1 e^{r_1 \delta t} + A_2 e^{r_2 \delta t} + d/c \\
    T_{m, t+1} =& A_1A_3 e^{r_t \delta t} + A_2A_4 e^{r_2 \delta_t} + d/c.
\end{align*}

\subsubsection{Solar gain}
\label{sec:app_solargain}
It is possible to add the solar gain to the simulator. It is computed based on the CIBSE Environmental Design Guide \cite{CIBSE}.

The house's lighting characteristics $\theta_S$, which include the window area and the shading coefficient of 0.67 are needed to model the solar gain, $Q_{s,t}$.

Then, the following assumptions are made:
\begin{itemize}
    \item The latitude is 30$^\circ$.
    \item The solar gain is negligible before 7:30 am and after 5:30 pm at such latitude. 
    \item The windows are distributed evenly around the building, in the 4 orientations.
    \item All windows are vertical.
\end{itemize}

This allows us to compute the coefficients of a fourth-degree bivariate polynomial to model the solar gain of the house based on the time of the day and the day of the year. 

\subsection{Detailed air conditioner model}
\label{sec:app_air_conditioner_model}

Once again based on the Gridlab-D Residential module user's guide \cite{GridLabD}, we model the air conditioner's power consumption $P_{a,t}$ when turned \textsc{on}, and the heat retrieved from the air $Q_{a,t}$, based on its characteristics $\theta_H$, such as cooling capacity $K_a$, coefficient of performance $COP_a$, and the latent cooling fraction $L_a$.

$COP_a$ and $L_a$ are considered constant and based on default values of the guide: $COP_a = 2.5$ and $L_a = 0.35$.
We have:
\begin{align*}
    Q_{a,t} &= - \dfrac{K_a}{1+L_a} \\
    P_{a,t} & = \dfrac{K_a}{COP_a}.
\end{align*}
We set $K_a$ to 15 kW, or 50 000 BTU/hr, to be able to control the air temperature even with high outdoor temperatures. This is higher than most house ACs, but allows to have sufficient flexibility even at high outdoor temperatures (a 5kW AC would have to be always \textsc{on} to keep a 20$\degree$C temperature when it is 38$\degree$C outside). This choice does not significantly affect our results: with lower outdoor temperatures, the problem is equivalent with lower AC power.

\subsection{Regulation signal}
\label{sec:app_reg_signal}

\subsubsection{Interpolation for the base signal}
As described in Section \ref{sec:reg_signal}, we estimate $D_{a,t}$ by interpolation. A bang-bang controller is ran without lockout for 5 minutes, and we compute the average power that was consumed. This gives a proxy for the amount of power necessary in a given situation.

A database was created by estimating $D_{a,t}$ for a single house for more than 4 million combinations of the following parameters: the house thermal characteristics $\theta_h$, the differences between its air and mass temperatures $T_{a,t}$ and $T_{m,t}$ and the target temperature $T_T$, the outdoor temperature $T_{o,t}$, and the AC's cooling capacity $K_a$. If the solar gain is added to the simulation, the hour of the day and the day of the year are also considered.

When the environment is simulated, every 5 minutes, $D_{a,t}$ is computed by summing the interpolated necessary consumption of every house of the cluster. 
The interpolation process is linear for most parameters except for the 4 elements of $\theta_h$ and for $K_a$, which are instead using nearest neighbours to reduce the complexity of the operation.

\subsubsection{Perlin noise}
\label{sec:app_perlin}
1-D Perlin noise is used to compute $\delta_{\Pi,t}$, the power generation high-frequency element. Designed for the field of computer image generation, this noise has several interesting properties for our use case. 

Perlin noise is most of the time generated by the superposition of several sub-noises called octaves. It is possible to restrict the span of the values that they can take. Thus, it is possible to test the agents in an environment taking into account several frequencies of non-regular noise, but whose values are restricted within realistic limits. Moreover, the average value of the noise can be easily defined and does not deviate, which ensures that for a sufficiently long time horizon, the noise average is 0. 

Each octave is characterized by 2 parameters: an amplitude and a frequency ratio. The frequency represents the distance between two random deviations. The amplitude represents the magnitude of the variation. Normally the frequency increases as the amplitude decreases. This way, high-amplitude noise is spread over a wider interval and lower amplitude noise is more frequent and compact. 

\begin{figure*}
    \centering
    \includegraphics[width=\textwidth]{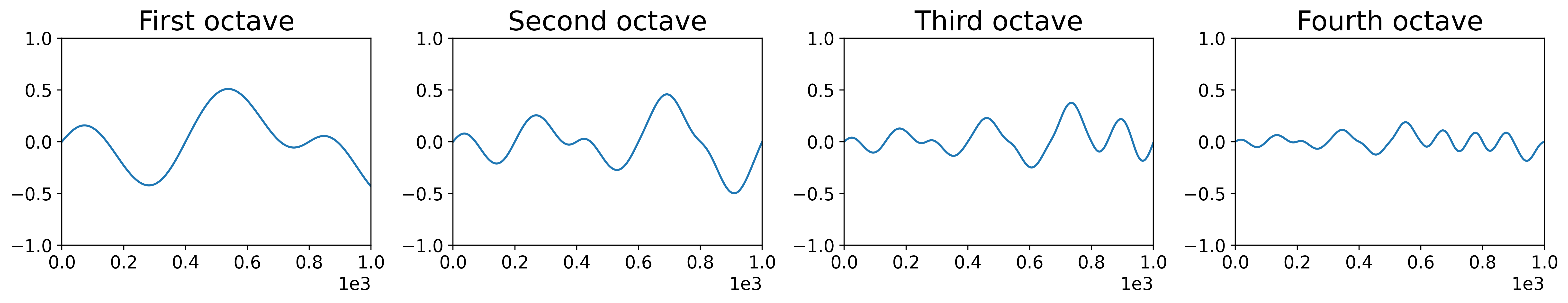}
    \includegraphics[width=150px]{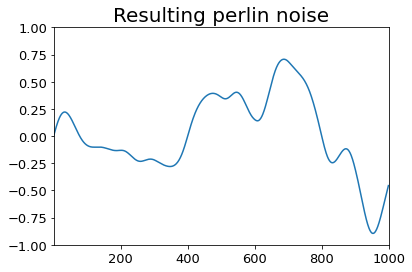}
    \caption{Illustration of how several octaves add up to form Perlin noise. The frequency of the octaves increases as their amplitude decreases.}
    \label{fig:perlinNoise}
\end{figure*}

In our case, we use 5 octaves, with an amplitude ratio of 0.9 between each octave and a frequency proportional to the number of the octave. 

%% file: Contents/Appendix/D_AlgorithmsDetails.tex
\section{Algorithm details}
\label{sec:app_algodetails}

\subsection{Model Predictive Control}
\label{sec:app_MPC}

Our MPC is based on a centralized model. At each time step, information about the state of the agents is used to find the future controls that minimize the reward function over the next $H$ time steps. The optimal immediate action is then communicated to the agents. At each time step, the algorithm calculates the ideal control combination for the $H$-time step horizon.

The cost function for both the signal and the temperature to minimize being the RMSE, the problem is modeled as a quadratic mixed-integer program. The solver used to solve the MPC is the commercial solver Gurobi \cite{gurobi} together with CVXPY \cite{cvxpy}. Gurobi being a licensed solver, its exact internal behavior is unknown to us and it acts as a black box for our MPC. However, we know that it solves convex integer problems using the branch and bound algorithm. The speed of resolution depends mainly on the quality of the solver's heuristics. 

The computation time required for each step of the MPC increases drastically with the number of agents and/or $H$. To be able to test this approach with enough agents and a rolling horizon allowing to have reasonable performance, it was necessary to increase the time step at which the agents make decisions to 12 seconds (instead of 4 for other agents).

It was impossible to launch an experiment with the MPC agent for 48 hours in a reasonable time. To compensate, we launched in parallel 200 agents having been started at random simulated times. In order to reach quickly the stability of the environment, the noise on the temperature was reduced to 0.05$\degree$C. We then measured the average RMSE over the first 2 hours of simulation for each agent.

Despite this, it was impossible to test the MPC with more than 10 agents while keeping the computation time reasonable enough to be used in real time. That is to say, in a time shorter than the duration between two-time steps.

At each time step, the MPC solves the following optimization problem :
\begin{equation*}
\min_{a \in \left\{0,1\right\}^{N \times H}} \sum_{t \in H}\alpha_{\mathrm{sig}} (\sum_{i \epsilon N} P_{i,t} - s_0)^2 + \alpha_{\mathrm{temp}} \sum_{i \epsilon N} (T_{h,t,i} -T_{t,t,i})^2,
\end{equation*}
such that it obeys the following physical constraints of the environment:
\begin{align*}
    T_{h,t,i}, T_{m,t,i} = F_1(a_{i,t}, T_{h,t-1,i},T_{m,t-1,i}) \ & \forall \ t \in H, i \in N \\
    P_{h,t,i} = a_{i,t} F_2(\theta_{a}^i)  \  & \forall \ t \in H, i \in N,
\end{align*}
and the lockout constraint:
\begin{equation*}
    l_{max}(a_{i,t} - \omega_{i,t-1}) - \sum_{k=0}^{l_{max}} (1-\omega_{i,t-k})\leq 0 \ \forall \ t \in H, i \in N,
\end{equation*}
where $F_1$ and $F_2$ are convex functions that can be deduced from the physical equations given in Section \ref{sec:app_environmentdetails}.

\subsection{Learning-based methods}
\label{sec:app_learning_based}

\subsubsection{TarMAC and MA-PPO}
\label{sec:App_TarMAC}
The original implementation of TarMAC \cite{Das_2019} is built over the Asynchronous Advantage Actor-Critic (A3C) algorithm \cite{A3C}. The environments on which it is trained have very short episodes, making it possible for the agents to train online over the whole memory as one mini-batch.

This is not possible with our environment where training episodes last around 16000 time steps. As a result, we built TarMAC over our existing MA-PPO implementation. The same loss functions were used to train the actor and the critic.

The critic is given all agents' observation as an input.

\begin{figure*}
    \centering
    \includegraphics[width=\textwidth]{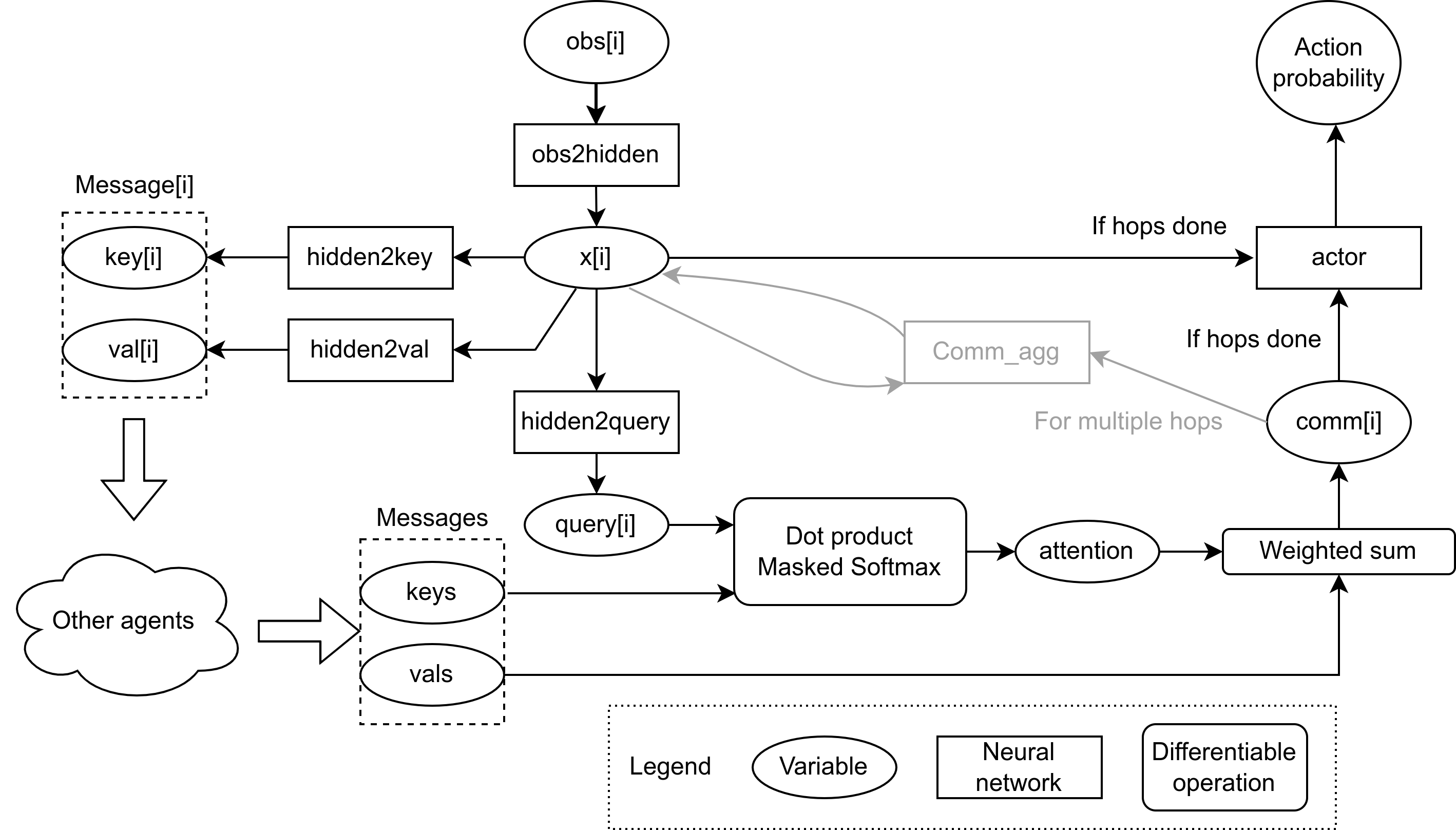}
    \caption{Architecture of the TarMAC-PPO actor}
    \label{fig:ppo_actor}
\end{figure*}

The actor's architecture is described in Figure~\ref{fig:ppo_actor}.
Agent $i$'s observations are passed through a first multi-layer perceptron (MLP), outputting a hidden state $x$. $x$ is then used to produce a key, a value, and a query by three MLPs. The key and value are sent to the other agents, while agent $i$ receives the other agents' keys and values. The other agents' keys are multiplied using a dot product with agent $i$'s query, and passed through a softmax to produce the attention. Here, a mask is applied to impose the localized communication constraints and ensure agent $i$ only listen to its neighbours. The attention is then used as weights for the values, which are summed together to produce the communication vector for agent $i$. For multi-round communication, the communication vector and $x$ are concatenated and passed through another MLP to produce a new $x$, and the communication process is repeated for the number of communication hops. Once done, the final $x$ and communication vector are once more concatenated and passed through the last MLP, the actor, to produce the action probabilities.

We take advantage of the centralized training approach to connect the agents' communications in the computational graph during training. Once trained, the agents can be deployed in a decentralized way.

\subsubsection{Neural networks architecture and optimization}
For MA-DQN as well as for MA-PPO-HE, every neural network has the same structure, except for the number of inputs and outputs. The networks are composed of 2 hidden layers of 100 neurons, activated with ReLU, and are trained with Adam \cite{Kingma_Ba_2017}.

For TarMAC-PPO, the actor's \textit{obs2hidden}, \textit{hidden2key}, \textit{hidden2val}, \textit{hidden2query} and \textit{actor} MLPs (as shown in Figure \ref{fig:ppo_actor}) all have one hidden layer of size 32. \textit{obs2hidden} and \textit{actor} are activated by ReLU whereas the three communication MLPs are activated by hyperbolic tangent. The hidden state $x$ also has a size of 32. 

The centralized critic is an MLP with two hidden layers of size 128 activated with ReLU. The input size is the number of agents multiplied by their observation size, and the output size is the number of agents.

For all networks, the inputs are normalized by constants to facilitate the training. The networks are optimized using Adam.

\subsubsection{Hyperparameters}
We carefully tuned the hyperparameters through grid searches. Table~\ref{tab:hyperparameters} shows the hyperparameters selected for the agents presented in the paper.

\begin{table}[h]
    \centering
    \caption{Training hyperparameters}
    \begin{tabular}{|c|c|c|c|}
        \hline
        Hyperparameter & TarMAC-PPO & MA-PPO & DQN  \\
        \hline
        Learning rate & 0.001 & 0.001 & 0.0001 \\
        Mini-batch size & 256 & 512 & 256 \\
        Clip parameter & 0.2 & 0.2 & - \\
        Max grad norm & 0.5 & 0.5 & - \\
        Number epochs & 200 & 200 & - \\
        Number updates & 10 & 10 & - \\
        Number episodes & 200 & 200 & - \\
        Discount factor $\gamma$ & 0.99 & 0.99 & 0.99 \\
        Key vector size & 4 or 8 & - & - \\
        Comm. vector size & 8 & - & - \\
        Number comm. rounds & 1 & - & - \\
        Buffer capacity & - & - & 65536 \\
        $\epsilon$ decay & - & - & 0.995 \\
        Min $\epsilon$ & - & - & 0.01 \\
        
        \hline
    \end{tabular}
    \label{tab:hyperparameters}
\end{table}

%% file: Contents/Appendix/E_Proofs.tex
\section{$N_\mathrm{de}$ and per-agent RMSE}
\label{sec:app_proof_smallererror}

In this section, we discuss the relation between the per-agent signal RMSE of an aggregation of $N$ homogeneous agents if $N$ is multiplied by an integer $k \in \mathbb{N}$.

We consider the aggregation of size $kN$ as the aggregation of $k$ homogeneous groups $g_j$ of $N$ agents which consumes a power $P^j_{g,t} = \sum_i^N P^i_{a,t} $. We have: $P_t =  \sum_i^{kN} P^i_{a,t} = \sum_j^k P^j_{g,t}$. 

We assume that each group tracks an equal portion of the signal $s^j_t =  s_t/k$. We assume that the tracking error $P^j_{g,t} - s^j_t$ follows a 0-mean Gaussian of standard deviation $\sigma_g$. This Gaussian error is uncorrelated to the noise of other groups.

It follows from the properties of Gaussian random variables that the aggregation signal error $P_t - s_t$ follows a Gaussian distribution of mean $\mu_k = 0$ and standard deviation $\sigma_k = \sqrt{k}\sigma_g$ for all $k \geq 1$ with $k \in \mathbb{N}$.

Hence the signal's RMSE of a group of $kN$ agents, which is a measured estimation of $\sigma_k$, is approximately $\sqrt{k}$ times the RMSE of a group of $N$ agents, which estimates $\sigma_g$. Finally, the per-agent RMSE is computed as the group's RMSE divided by the number of agents. We therefore have that the per-agent RMSE of $kN$ agents is approximately $\sqrt{k}/k = 1/\sqrt{k}$ times the RMSE of $N$ agents.

This discussion provides an intuitive explanation for the diminution of the relative RMSE when the number of agents increases. However, it is based on the assumption that the error of each group is not biased, which is not necessarily true with our agents. This explains why the RMSEs are not 10 times lower passing from $N_\mathrm{de} = 10$ to $N_\mathrm{de} = 1000$.